\documentclass[journal]{IEEEtran}

\ifCLASSINFOpdf
  \usepackage[pdftex]{graphicx}
  \graphicspath{{../pdf/}{../jpeg/}}
  \DeclareGraphicsExtensions{.pdf,.jpeg,.png}
\else
\fi

\def\pprw{8.5in}
\def\pprh{11in}

\usepackage[utf8]{inputenc}
\usepackage{cite}
\usepackage{amsfonts}

\usepackage{amsthm}
\usepackage[super]{nth}

\usepackage{multirow}
\usepackage{footnote}
\usepackage{xcolor}
\usepackage{colortbl}
\usepackage{stfloats}
\usepackage[hyphens]{url}
\usepackage{subfigure}
\usepackage{lipsum} 
\usepackage{fancyhdr}
\usepackage{amsmath}
\usepackage{algorithmic}
\usepackage{caption}
\usepackage{comment}
\usepackage{graphicx}
\usepackage[outdir=./figures/]{epstopdf}

\begin{document}
%
\title{Large-scale analysis of user exposure to online advertising in Facebook}
%
%
%


\author{\IEEEauthorblockN{Aritz Arrate,
Jos\'e Gonz\'alez Caba\~nas, \'Angel Cuevas, Mar\'ia Calder\'on and Rub\'en Cuevas\\Department of Telematic Engineering,
Universidad Carlos III de Madrid}}

\maketitle

\begin{abstract}
Online advertising is the major source of income for a large portion of Internet Services. There exists a body of literature aiming at optimizing ads engagement, understanding the privacy and ethical implications of online advertising, etc. However, to the best of our knowledge, no previous work analyses at large scale the exposure of real users to online advertising. This paper performs a comprehensive analysis of the exposure of users to ads and advertisers using a dataset including more than 7M ads from 140K unique advertisers delivered to more than 5K users that was collected between October 2016 and May 2018. The study focuses on Facebook, which is the second largest advertising platform only to Google in terms of revenue, and accounts for more than 2.2B monthly active users. Our analysis reveals that Facebook users are exposed (in median) to 70 ads per week, which come from 12 advertisers.  Ads represent between 10\% and 15\% of all the information received in users' newsfeed. A small increment of 1\% in the portion of ads in the newsfeed could roughly represent a revenue increase of 8.17M USD per week for Facebook.  Finally, we also reveal that Facebook users are overprofiled since in the best case only 22.76\% of the active interests Facebook assigns to users for advertising purpose are actually related to the ads those users receive.

\end{abstract}

\begin{IEEEkeywords}
Facebook, online advertising, transparency, user-centric analysis, human-computer interaction.
\end{IEEEkeywords}

%
\IEEEpeerreviewmaketitle

\section{Introduction}
\label{sec:intro}
The Interactive Advertising Bureau (IAB) reported that the online advertising revenue was \$88B in 2017 only in US, which represents a yearly growth of 21.4\% from 2016 \cite{IAB17}. Online advertising represents the most important source of income for online services and websites in the Internet. For instance, online advertising represents more than 85\% of the revenue for some of the largest Internet companies like Google \cite{GoogleinvestorsQ22018} or Facebook (FB) \cite{FBinvestorsQ22018}. Therefore, the Internet sustainability is currently linked to the health of the online advertising ecosystem.

The great importance of the online advertising ecosystem in the current Internet has motivated researchers to investigate in this area. There is a body of the literature that aims at improving online advertising from different angles. Some works propose solutions to improve the efficiency of digital marketing through online advertising to maximize the benefits of online services \cite{intro_perf1}\cite{intro_perf2}\cite{intro_perf3}\cite{howTarget}. Other research papers analyze users' perception of online advertising and propose what are the best strategies to achieve a positive perception \cite{intro_userpercep1}\cite{intro_userpercep2}\cite{intro_userpercep3}\cite{intro_userpercep4}\cite{intro_userpercep5}. Other part of the literature aims at creating transparency and protect users privacy in an ecosystem that many times collects, process and (indirectly) offers Internet users personal data to increase the revenue of different stakeholders \cite{intro_privacy1}\cite{intro_privacy2}\cite{Jose_usenix} . 

Despite the literature in the area of online advertising, to the best of our knowledge there is no prior study that addresses the exposure of real end-users to online advertising at large-scale. We could only find some previous works that use bots  instead of real users to measure the exposure to online advertising using rather limited datasets. \cite{adscape}\cite{Juanmi_Conext}. 

Our work presents the first large-scale study which analyses the exposure of real Internet users to the online advertising ecosystem.  In this research we focus on one of the most important online advertising stakeholders, Facebook. The reasons why we choose FB for this research are: (i) Its online advertising platform is second only to Google in terms of revenue \cite{FB2nd}; (ii) FB implements itself a complete advertising ecosystem; (iii) FB is one of the most popular online services with more than 2B active users; (iv) there is no previous work that analyzes the exposure of FB users to online advertising in this platform, but just some few studies focusing on privacy and transparency of advertising in FB \cite{Jose_usenix}\cite{EURECOM+5420}\cite{speicherpotential}.

In particular, this work aims to address four fundamental elements regarding the exposure of FB users to online advertising in this platform. 

First, we quantify the exposure of users to ads. Our goal is to reveal how many ads users receive in different standard time windows (e.g., week), and measure which is the portion that those ads represent among all the received information by FB users in their newsfeed. Finally, we also analyze whether all users in FB are exposed to the same volume of ads or instead some users are actually exposed to a larger portion of ads.

Second, we measure the exposure of users to advertisers. Many times users receive more than one ad from the same advertiser, thus, knowing the exposure of users to ads does not allow to know whether they have been exposed to few or many advertisers. Our goal is to quantify how many advertisers reach users with ads in different time windows (session\footnote{A session comprehends the time that a user has had Facebook open in a browswer's tab, either in the foreground or background. It ends either if the user closes the Facebook session, or closes the tab.}, day, week and month), and what is the distribution of delivered ads across advertisers. This means whether few advertisers are responsible for a large portion of the ads delivered to a user, or rather ads are homogeneously distributed across advertisers. 

Third, we aim at understanding whether FB \textit{overprofiles} users or not. We analyze which portion of the interests (also known as ad preferences) FB assigns to users (to allow advertisers perform targeted advertising campaigns) is actually related to the ads they receive.  Our hypothesis is that most of those interests are irrelevant for advertisers, and thus, are not used to target users.

Fourth, we compute the engagement of FB users with the receive ads. We consider a user engaged with an ad when she clicks on the ad. We measure the probability that an advertiser has to engage a user according to the number of impressions the user receives from that advertiser. This will allow to understand whether overwhelming users with many ads is a successful strategy to engage a user or not. 

To accomplish this research work we rely on a large dataset collected between October 2016 and May 2018 that includes more than 7M ads delivered to more than 5K real users by 140K unique advertisers. To the best of our knowledge this is the largest dataset used in a study that analyzes the exposure of users to online advertising. 

To conclude the introduction we summarize the main findings of our research:

\noindent (1) In median FB users consume 70 ads per week, 6 ads per FB session and 0.8 ads per minute while browsing in FB. 

\noindent (2) Ads represent between 10\% and 15\% of the content consumed by FB users in their newsfeed; 

\noindent (3) Not all users are exposed to the same volume of ads. We found users that systematically receive more ads in their newsfeed per post of information than others. 

\noindent (4) In average, users are reached by 4, 12, 33 and 67 advertisers per session, day, week and month, respectively. 

\noindent (5) In median, 10 advertisers are responsible for half of the ads displayed to a FB user in the time span of our study. 

\noindent (6) FB users are overprofiled. That means they are tagged with many interests that have no relation with the ads they receive. In the best case (for Facebook) only 22.76\% of the interests FB assigns to the users are related to the ads they receive. 

\noindent (7) The probability that an advertiser engages a user (i.e., gets a click on an ad) grows logarithmically with the number of ads impressions the user receives from the advertiser. The first impression of the ad is the one presenting the highest probability to engage the user. After the first ad impression,  the probability of getting a click from the user is homogeneously distributed across the rest of impressions of the ad.

\section{Background}
In this section we briefly describe the FB online advertising platform and the FDVT tool that has been used to collect the data used in this paper.

\begin{figure}[t]
	\centering
	\includegraphics[width=\columnwidth]{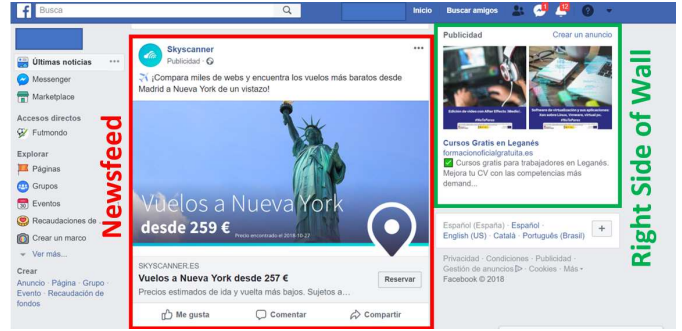}
	\caption{Snapshot of Facebook wall showing the two locations were ads are delivered in Facebook: newsfeed and right-side of the wall.. The red box highlights an ad delivered in the newsfeed of the user. The green box highlights an ad delivered in the right-side of the wall.}
	\label{fig:newsfeedDetail}
\end{figure}

\subsection{Facebook Online Advertising}
The last Facebook quarter report (Q3 2018) to investors \cite{FBinvestorsQ32018} shows that 98\% of its revenue comes from online advertising. FB offers an online advertising platform in which advertisers can launch micro-targeting campaigns to reach users with ads. 

Facebook profiles users with the so-called ad preferences, which are interests that may be relevant for users. Ad preferences are inferred from the activity of users in Facebook\footnote{The ad preferences assigned to the users, are also known as interests when the advertisers determine the target audience of a given campaign. For this reason, in this paper we will use indistinguishably the terms "interests" or "ad preferences".} and its third parties. 

Then, when an advertiser configures a campaign targeting users in \textit{France} interested in \textit{Soccer}, all \textit{French} FB users assigned the ad preference \textit{Soccer} are potential candidates to receive the ad associated to this campaign. 

To decide which ad is delivered to a user a real-time auction takes place. FB delivers the ad of the advertiser winning the auction. To participate in the auction advertisers have placed in advance their bids to display their ads, which is the price they are willing to pay for delivering its ad to users matching the audience they are targeting. Although there are several pricing models in the online advertising industry the most popular ones, also in FB, are: (i) \textit{Cost Per Mile (CPM)} in which the advertiser bid refers to the price it is willing to pay for delivering one thousand impressions of its ad to users matching the target audience; (ii) Cost Per Click (CPC) in which the advertiser bid refers to the price it is willing to pay for each click in the delivered ad from a user matching the target audience. The CPC and CPM values together with other parameters are the input to the auction algorithm of FB. This algorithm is an asset of the company and thus FB keeps it secret.

To conclude this brief overview of the FB advertising platform we describe where ads are delivered within the FB front-end. In the case of laptops/desktops, there are two places where ads are displayed: the newsfeed of the user and the right side of the wall. Figure \ref{fig:newsfeedDetail} depicts a snapshot highlighting in red an ad displayed in the newsfeed, and in green an ad displayed in the right side of the wall. It is important to note that ads in the newsfeed are only displayed when the user is browsing her own newsfeed, and hence, this type of ads are not delivered in the case the user browses walls of friends or FB pages. On the contrary, right-side ads are delivered at any time of the browsing. We note that this paper uses data exclusively collected from desktops/laptops. In the case of mobile devices the right-side space does not exist, thus ads are only delivered in the newsfeed.

\subsection{FDVT: Data Valuation Tool for Facebook users}

The FDVT \cite{FDVT} is a web-browser plug-in that informs Facebook users of an estimation of the revenue they are generating for FB based on the ads they receive while browsing in FB. It uses a small profile (country, gender, age and relationship status) to obtain in real-time the CPM and CPC prices advertisers have recently paid to reach users with that profile. In addition, the FDVT monitors the ads delivered to the user while browsing in FB. Using the CPM associated to the user profile and the number of ads delivered summed to the CPC and the number of clicks of the user on ads during a session, the FDVT computes in real-time an estimation of the revenue generated by that user. The FDVT has been installed by more than 7000 users since it was publicly released in October 2016. 

When installing the FDVT users grant permission to use the anonymous collected information for research purposes. Hence, this study is compliant with the current data protection regulation in Europe. The most relevant information collected by the FDVT in the context of this paper is the data related to the ads delivered to users. In particular, the FDVT collects: the timestamp when the ad is delivered, the position of the ad (either newsfeed or right side) and the url(s) embedded in the ad that forwards the user to the advertiser landing page associated to the ad in the case she clicks on the ad. In addition, the FDVT also collects the number of information posts different than ads delivered in the user's newsfeed. All this information is collected for each FB session of a user.

For a more detailed description of the FDVT we refer the reader to \cite{FDVT}.
\section{Dataset}
\label{sec:dataset}
To comprehensively cover the research questions addressed in this paper we base our study on a large dataset (see Table \ref{tab:dataset2} including more than 7M ads (embedding more than 20M URLs) delivered to 5,468 FDVT users across 589K sessions. This dataset was collected between October 2016 and May 2018. Note that the users that have installed the FDVT do not necessarily represent a random sample of FB users. In addition, we had to find who were the advertisers behind these ads. In this paper, we identify the advertisers by the web-domain associated to the landing page a user would reach if she clicks on the ad (e.g., adidas.com). Moreover, we categorized the advertisers in order to map the ads delivered to the ad preferences assigned to the users. Following, we describe each of this processes and present a descriptive analysis of the final dataset. 

\begin{table}[t]
	\centering
	\begin{tabular}{ccccc}
		\rowcolor[HTML]{343434} 
		{\color[HTML]{FFFFFF} URLs}& 
		{\color[HTML]{FFFFFF}Unique URLs}&
		{\color[HTML]{FFFFFF}Ads}&
		{\color[HTML]{FFFFFF}Sessions}&
		{\color[HTML]{FFFFFF}Users}\\ 
		\rowcolor[HTML]{FFFFFF} 
		{\color[HTML]{333333}20,908,626}    &   {\color[HTML]{333333}14,372,631} & {\color[HTML]{333333}7,546,740}  &   {\color[HTML]{333333}589,911}    &   {\color[HTML]{333333}5,468}   \\
	\end{tabular}
	\caption{Number of URLs, unique URLs, ads, sessions and users stored in the dataset used in this work.}
		\label{tab:dataset2}
\end{table}

\begin{table}[t]
	\centering
	\begin{tabular}{ccccc}
		\rowcolor[HTML]{343434} 
		{\color[HTML]{FFFFFF} }&
		{\color[HTML]{FFFFFF} Analyzable}&
		{\color[HTML]{FFFFFF} Advertiser Obtained}&
		{\color[HTML]{FFFFFF} Categorized}\\
		\rowcolor[HTML]{FFFFFF}
		\cellcolor[HTML]{656565}{\color[HTML]{FFFFFF}Number of Ads} &
		{7,339,059} &
		{5,420,863}   &
		{4,825,023}\\
	\end{tabular}
	\caption{Number of ads analyzed, number of advertisers identified through the ads landing page, and number of advertisers categorized using McAfee.}
	\label{table:analyzable}
\end{table}

\subsection{Method to obtain advertisers' identity}
Each ad in Facebook embeds one or more URLs that forward the user to the advertiser's landing page in case the user clicks on the ad. Overall, we analyzed more than 20M URLs from which more than 14M were unique. The way FB users are redirected to the landing page differs from URL to URL. We classified the URLs into four groups depending on the way they redirect users to the landing page: (1) URLs that redirect users to a domain different than FB but in which the redirection is not performed by FB (38,102 URLs), (2) URLs that redirect users to a domain different than FB in which the redirection is performed by FB (10,948,398  URLs), (3) internal Facebook URLs redirecting users to Facebook Pages (1,807,560 URLs), and (4) the remaining internal Facebook URLs (1,578,571 URLs). For the last group we could not find the advertiser because the landing page inside FB is not directly linked to any advertiser's website or name. For the remaining three groups, we implemented three different methods to reach the landing page and retrieve the advertiser name (i.e., main web-domain name). 

For groups (1) and (2), we implemented a simple Python module deactivating the Transport Layer Security (TLS) \cite{tlsrfc} requirement to obtain the Top Level Domain (TLD) of the landing page associated to a given URL.  We note that to reach the landing page we gathered all the intermediary domains that act as redirections to the landing page. The only variation in the methodology applied in groups (1) and (2) relays in establishing Facebook as the first redirection for those URLs in group (2). 

For group (3), we had to obtain the domains associated to FB pages. To that end we used a hybrid methodology to speed-up the collection process. On the one hand, we leveraged the FB graph API \footnote{Facebook Graph API: \url{https://developers.facebook.com/docs/graph-api/}} that provides the URL associated to public FB pages, and, on the other hand, we implemented a web-scraper that retrieves FB pages' domains. 

It is important to note that we failed on retrieving the advertiser associated to some of the ads for the two following reasons: (i) the period we collected the ads spans 20 months from October 2016 to May 2018, and the process to retrieve the advertisers behind those ads was implemented after May 2018. Hence, we found that some of the URLs were not leading anymore to a landing page; (ii) some times the chain of redirections got stuck in an intermediary (e.g., an ad-network) and we could not reach the final landing page.

As summarized in Table \ref{table:analyzable}, we overall analyzed more 7M ads and were able to identify their associated domain (i.e., amazon.com) for almost 75\% (5.4M) of them. 

\subsection{Advertiser classification}

Once we have retrieved the advertisers associated to the ads, the next step was to classify them assigning one or more categories related to their content (e.g., sports, online shopping, dating, job search, etc.). To this end we used an online service offered by McAfee \cite{McAfee} which classifies web domains using 109 different categories\footnote{TrustedSource Web Database Reference Guide (Category Set 4) \url{https://kc.mcafee.com/corporate/index?page=content&id=PD22571}}. This service allows users to introduce a web domain name and returns the categories assigned to that website. We implemented an automated software that leveraged McAffee's classification service requesting one by one the categories of the 5.4M web domains associated to the ads for which we were able to find the associated landing page.

We categorized the advertisers associated to 4.8M ads (87\%) among all the ads for which we retrieved the advertiser (see table \ref{table:analyzable}).  

\begin{figure*}[t]
    \centering
	\subfigure[Total ads]{\includegraphics[scale=0.23]{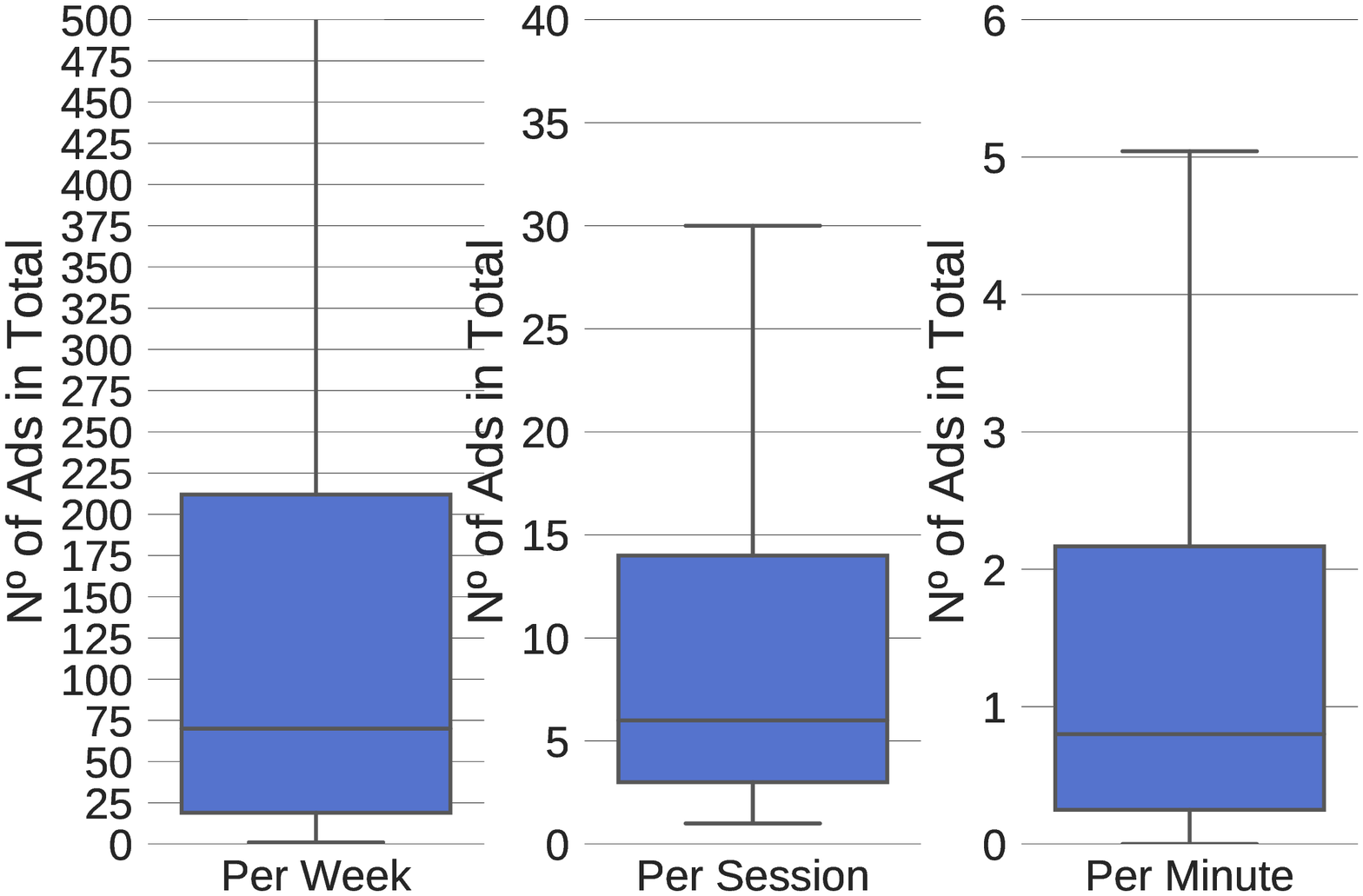}\label{fig:num_ads_total}}
	\subfigure[Ads newsfeed]{\includegraphics[scale=0.23]{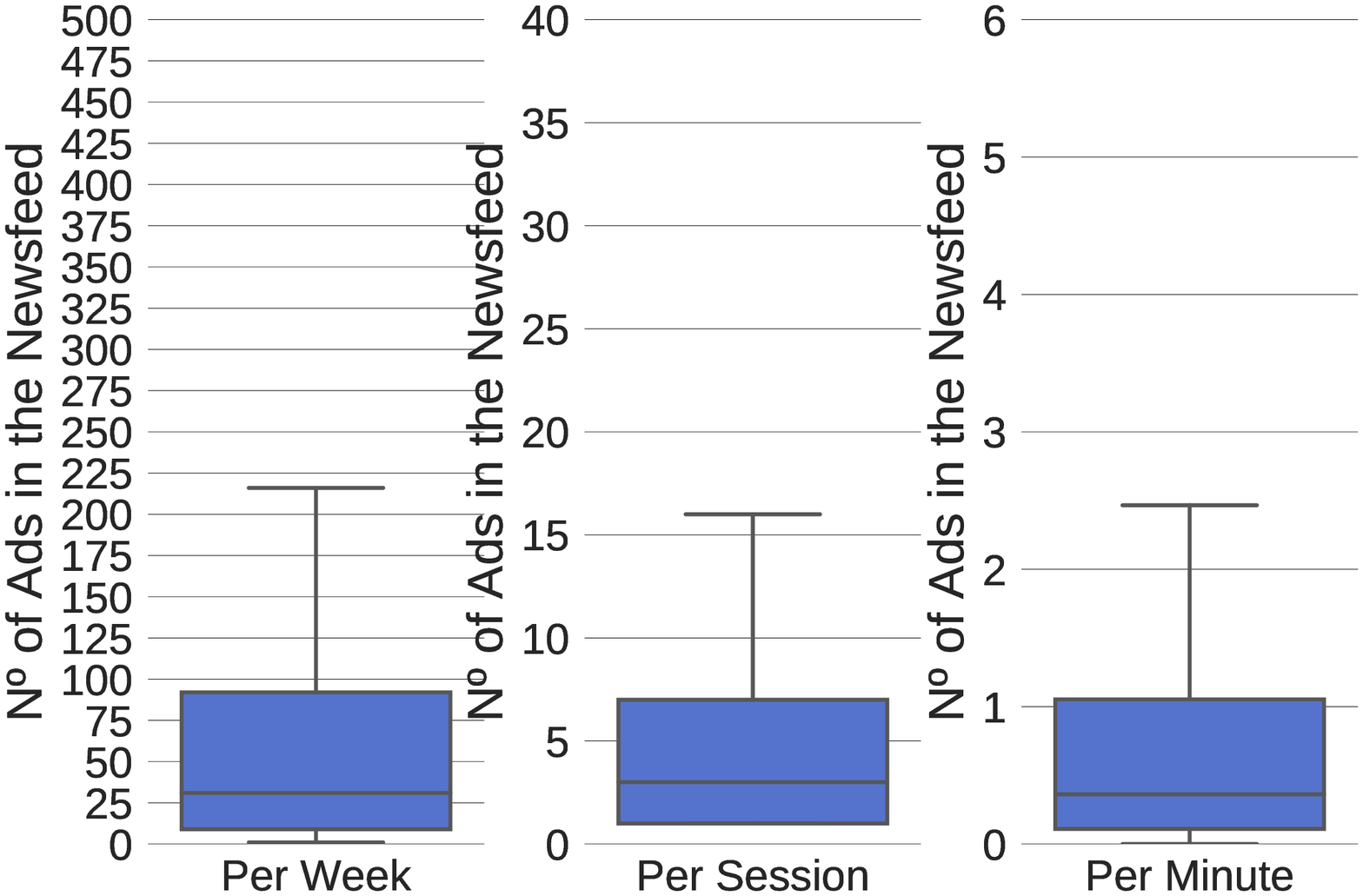}\label{fig:num_ads_news}}
	\subfigure[Ads right]{\includegraphics[scale=0.23]{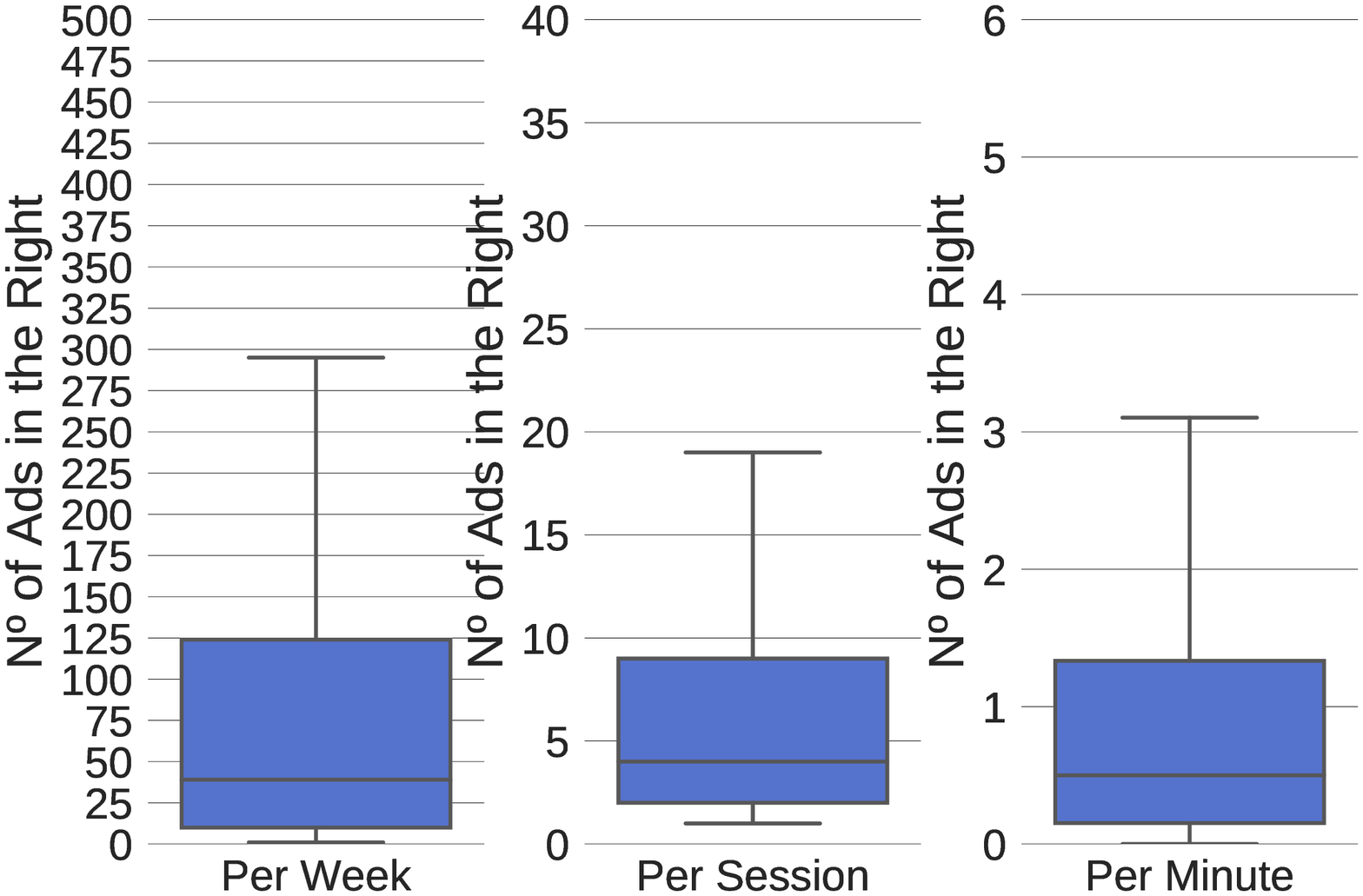}\label{fig:num_ads_right}}
	\caption{Boxplot showing the distribution of the number of ads received by FDVT users in a week, in a session and in a minute.}
	\label{fig:distributionAdsGeneral}
\end{figure*}

\begin{figure*}[t]
	\centering
	\includegraphics[width=\textwidth]{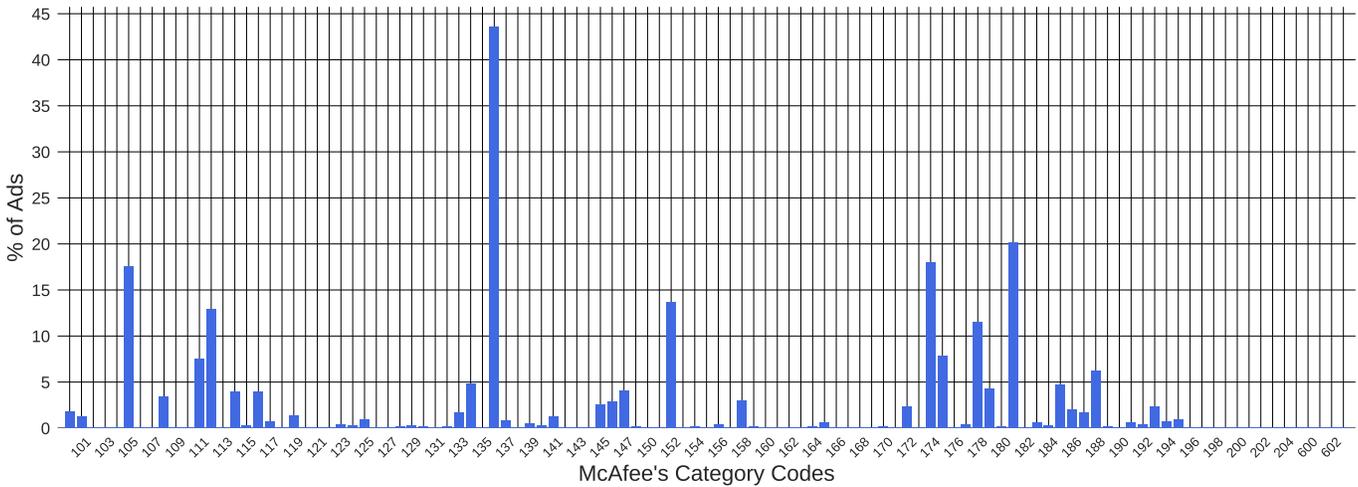}
	\caption{Histogram of the portion of ads associated to each Mcafee category assigned to advertisers landing web domains. Note that since Mcafee may assign up to three categories to a web domain the sum of the all the bars of the histogram is higher than 100\%.}
	\label{fig:distribution_categories}
\end{figure*}

\subsection{Dataset characterization}

We present a descriptive analysis of the dataset we will use to answer the research questions addressed in this paper. First, we quantify the volume of ads that users receive over time. Second, we evaluate whether the ads received by FDVT users are homogeneously distributed across the 109 McAfee categories or instead FDVT users are more exposed to ads from few categories.

Figure \ref{fig:distributionAdsGeneral} shows the distribution (in form of boxplot) of ads delivered to FDVT users considering three different time windows: one week, one session and one minute. In addition, we split the results considering all the ads received by a user (Figure \ref{fig:num_ads_total}), the ads delivered only in the newsfeed (Figure \ref{fig:num_ads_news}) and the ads received in the right side of the wall (Figure \ref{fig:num_ads_right}). 

In median a FDVT user receives 70 ads per week, 6 ads per FB session and 0.8 ads per minute while browsing in FB. Users receive in median 25\% more ads in the right side (39) than in the news feed (31) during a week. This is probably due to the fact that newsfeed ads only appear when the user is browsing its own wall, but they do not appear when she browses the walls of friends or external FB pages. Instead, right-side ads may appear when the user browses in friends walls or FB pages different than its own feed.     

Figure \ref{fig:distribution_categories} shows a histogram of the portion of ads belonging to each of the 109 McAffee categories used to classify websites. The results show that, as it could be expected, there are categories that are much more frequent than others among FB ads. The Top 10 categories are present in 71.02\% of the ads displayed to FDVT users. Those categories are: Online Shopping (present in 43.53\% of the ads), Marketing/Merchandising (20.1\%), Fashion/Beauty (17.93\%), Business (17.54\%), Travel (13.62\%), Entertainment (12.86\%), Internet Services (11.48\%), Software/Hardware (7.82\%), Education/Reference (7.47\%), Blogs/Wiki (6.25\%). Note that the sum of the probability of appearance of the categories is more than 100\% since McAfee may assign up to three categories to a website. For instance, the web domain associated to a perfume ad may be assigned the three top categories in the list, i.e., Online Shopping, Marketing/Merchandising and Fashion/Beauty.
\section{Exposure of users to ads}
\begin{figure}[t]
	\centering
	\includegraphics[width=\linewidth]{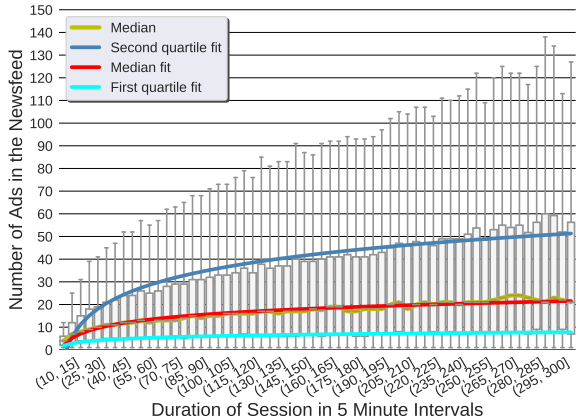}
	\caption{Distribution of the number of ads received by a user according to the duration of a session. The session time is represented in the x axis in bins of five minutes. Each five minutes bin includes a boxplot that shows the distribution of the number of ads for all FDVT sessions within that bin. The figure also includes the fit of the lower quartile, median and upper quartile.}
	\label{fig:distributionAdsFit}
\end{figure}

This section analyses and quantifies different aspects of the exposure of users to ads in FB to answer fundamental questions such as whether users are exposed to a large volume of ads or not, or if FB exposes all the users to the same volume of ads. 

\subsection{Ad exposure with respect to session time}
The first question we address is whether users' exposure to ads is linear over time,  that means if the number of ads a FB user receives increases linearly with the session time. Figure \ref{fig:distributionAdsFit} shows the number of ads received by FDVT users according to the duration of a session. The duration of the sessions is grouped into bins of 5 minutes, and for each bin figure \ref{fig:distributionAdsFit} depicts a boxplot showing the distribution of ads displayed per session including all the sessions within that bin.  The figure also shows the fitting of the boxplots \nth{25} percentile (lower-quartile), \nth{50} percentile (median) and \nth{75} percentile (upper-quartile). The logarithmic fitting is the one that minimizes the Mean Square Error (MSE). 

The number of ads received by FDVT users grows logarithmically with the session duration. This means that as the session time increases users are exposed to less ads. Our hypothesis is that the logarithmic behaviour observed happens because many FB sessions (especially the long ones) are linked to active and inactive periods of the users in FB. That means, users keep the session active and do not interact constantly with FB (e.g., they are browsing in other websites using a browser tab while FB is still open in a different tab), but come back to it from time to time.

\begin{figure*}[t]
	\centering
	\includegraphics[width=\linewidth]{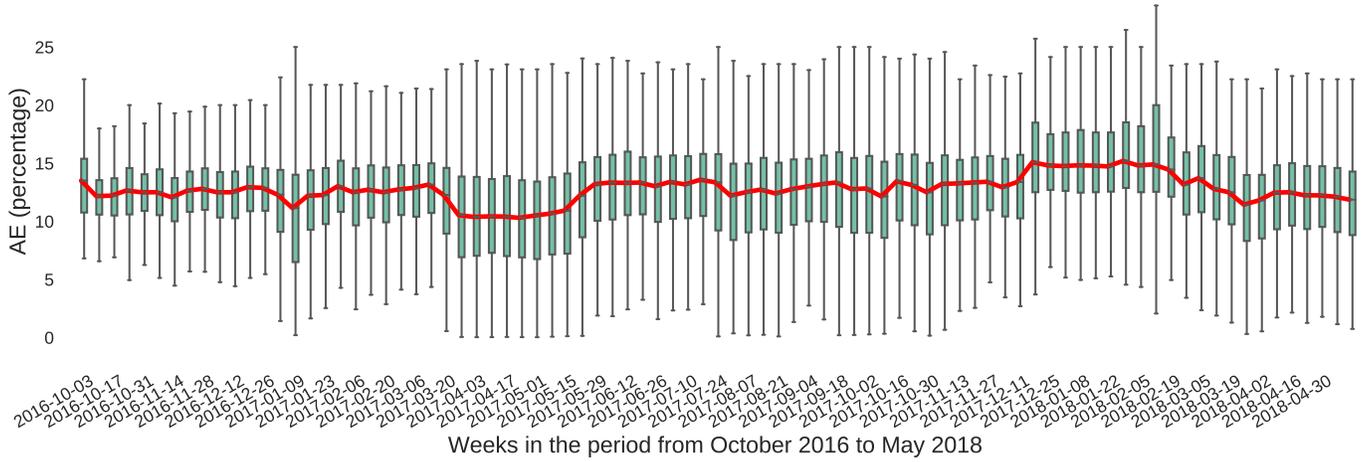}
	\caption{Evolution of the metric $AE =\frac{\#ads}{\#posts+\#ads}$ in the period October 2016 - May 2018 per week. Each week presents a boxplot of the metric that includes the distribution of the $AE$ across all the users with active sessions in each week. Note $AE$ is represented as percentage rather than portion in this figure.}
	\label{fig:distributionAdsTimePost}
\end{figure*}

\subsection{Ad exposure with respect to volume of posts}

The following question we address is what is the portion of ads users are exposed to in their newsfeed compared to the total volume of information including ads but more importantly information posts (e.g., posts from friends, FB pages, news outlets, etc). Answering this question helps to understand whether users are exposed to many ads or not. For instance, if 50\% of all the posts a user receives in her newsfeed were actually ads, we would conclude the exposure is very high since half of the information this user is consuming are ads. Instead, if the results showed that the exposure is only 1\% we would conclude that it is rather low. To answer the referred question we divide the number of ads displayed to a user in a FB session (i.e., \#ads)  by the total number of information posts (i.e., \#posts) plus ads the user has received in that session. We refer to this metric as $AE$ (Ad Exposure): $AE =\frac{\#ads}{\#posts+\#ads}$.

Figure \ref{fig:distributionAdsTimePost} depicts the $AE$ distribution per user and week between October 2016 and April 2018. Each week is represented with a box plot that denotes the distribution of the $AE$ across users. The goal of performing this experiment over time is to discover whether there are relevant variations of the $AE$ over time. The figure shows that in median the $AE$ ranges between 10\% and 15\% and does not show large variations over short periods of time. The period showing the lowest $AE$ value is April 2017 in which it peaks down to 10.29\% (1 ad per 8.71 information posts). In contrast,  the period with the highest $AE$ ranges between mid December 2017 and Mid February 2018 where the metric scales up to 15.19\% (1 ad per 5.58 information posts). The difference between the minimum and maximum $AE$ values in the analyzed period is translated to an exposure of 6.44 ads more per 100 information posts. Although, this difference may look small it might represent an important increase of the revenue obtained by FB. We have computed that an increase in the $AE$ of only 1\% from 12.5\% (the median value over time) to 13.5\% may roughly generate \$8.17M extra revenue per week for FB. To estimate the extra revenue we use: (i) the extra ads consumed in a week due to the 1\% $AE$ increase using as reference the median number of posts per week, which is 248; (ii) the median worldwide CPM in FB as of 23 October 2018, which was \$1.29; and, (iii) the number of active FB users that according to the 2018 Q3 investors report \cite{FBinvestorsQ32018} was 2.27B. The extra-revenue estimation would increase up to \$45.83M per week if the $AE$ increment evaluated was 4.90\%, which is the difference between the maximum (15.19\%) and minimum (10.29\%) $AE$ value observed in the period under analysis. Note these are just rough estimations that aim to show that even small tweaks in the $AE$ metric may lead to an important revenue increase for FB.       


\begin{figure}[htb!]
	\centering
	\includegraphics[width=\columnwidth]{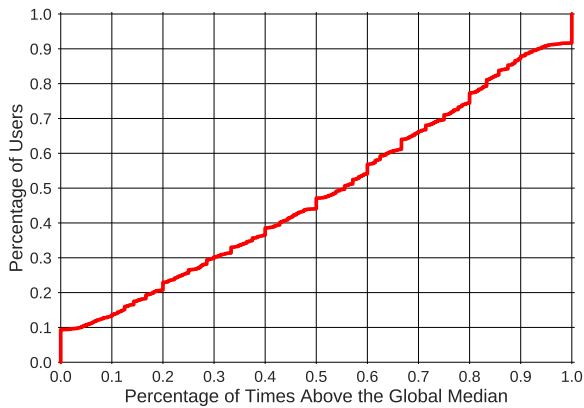}
	\caption{CDF representing the portion of times that a user is showing a weekly $AE$ value greater than the median $AE$ of the corresponding week in the period October 2016 - May 2018.}
	\label{fig:distributionTopUsers}
\end{figure}

\subsection{Comparison of ad exposure across users}

An immediate conclusion derived from observing Figure \ref{fig:distributionAdsTimePost} is that different users are exposed to a different $AE$. For instance, if we consider the first week in the figure starting on Oct. \nth{3}, 2016 the box plot denotes that for 1/4 of the FDVT users with active sessions that week the $AE$ was lower than 10.6\%, whereas for another 1/4 of the users was higher than 13.24\%. We wondered whether we could identify users for which the $AE$ is systematically above/below the median. That means whether there are FB users exposed to a larger fraction of ads in their newsfeed than others.

Figure \ref{fig:distributionTopUsers} shows a CDF that quantifies how frequently (x axis) users are above the weekly median $AE$. For instance the point \{x=0.55,y=0.5\} indicates that 50\% of the users have been above the median at most 55\% of the weeks in which they had active sessions. The linear shape of the CDF denotes a rather large diversity across users. This basically means that some FB users are exposed to more volume of ads given a fix volume of information posts. If all the users had a similar weekly $AE$, we would instead observe a vertical line around the x-axis value 0.5. The closer the CDF was to that vertical line the more similar would be the $AE$ across FB users, thus FB would be exposing users to a similar amount of ads with respect to the actual information they consume. Interestingly, 9.43\% of the users are always below the median, whereas 8.37\% of the users have been always above the median. This means that there exist users who are always exposed to a higher volume of advertising when browsing in Facebook.

\subsection{Summary of results}
The main outcomes of the analysis carried out in this section are: (i) the number of ads consumed by FB users grows logarithmically with the duration of the session; (ii) the portion of ads users consume in their newsfeed  ranges between 10\% and 15\%. We have proved that a slight increase of this metric may imply an important increment of FB revenue; (iii) there are FB users (8.37\% according to our analysis) systematically exposed to more ads compared to the rest of the users.

\section{Exposure of users to advertisers}
\label{sec:advertisers}

\begin{table}[t]
	\centering
	\begin{tabular}{cccccc}
		\rowcolor[HTML]{343434} 
		{\color[HTML]{FFFFFF} Metric}                       & {\color[HTML]{FFFFFF} General} & \multicolumn{1}{l}{\cellcolor[HTML]{343434}{\color[HTML]{FFFFFF} Monthly}} & \multicolumn{1}{l}{\cellcolor[HTML]{343434}{\color[HTML]{FFFFFF} Weekly}} &
		\multicolumn{1}{l}{\cellcolor[HTML]{343434}{\color[HTML]{FFFFFF} Daily}} &\multicolumn{1}{l}{\cellcolor[HTML]{343434}{\color[HTML]{FFFFFF} Session}} \\
		\rowcolor[HTML]{FFFFFF} 
		\cellcolor[HTML]{656565}{\color[HTML]{FFFFFF} mean} & {\color[HTML]{333333} 225.05}       & {\color[HTML]{333333} 67.37}                                                   & {\color[HTML]{333333} 32.87}                                                  & {\color[HTML]{333333} 12.22}& {\color[HTML]{333333} 4.52}                                                   \\
		\rowcolor[HTML]{C0C0C0} 
		\cellcolor[HTML]{656565}{\color[HTML]{FFFFFF} std}  & {\color[HTML]{333333} 290.11}       & {\color[HTML]{333333} 69.71}                                                   & {\color[HTML]{333333} 33.23}                                                  & {\color[HTML]{333333}11.28}&
		{\color[HTML]{333333} 3.52}                                                   \\
		\rowcolor[HTML]{FFFFFF} 
		\cellcolor[HTML]{656565}{\color[HTML]{FFFFFF} min}  & {\color[HTML]{333333} 1.00}       & {\color[HTML]{333333} 1.00}                                                   & {\color[HTML]{333333} 1.00}                                                  &
		{\color[HTML]{333333} 1.00}&{\color[HTML]{333333} 1.00}       \\
		\rowcolor[HTML]{C0C0C0} 
		\cellcolor[HTML]{656565}{\color[HTML]{FFFFFF} 25\%}  & {\color[HTML]{333333} 42.00}       & {\color[HTML]{333333} 18.00}                                                   & {\color[HTML]{333333} 9.00}                                                  &
		{\color[HTML]{333333} 5.00}&{\color[HTML]{333333} 3.00} \\
		\rowcolor[HTML]{FFFFFF} 
		\cellcolor[HTML]{656565}{\color[HTML]{FFFFFF} 50\%}  & {\color[HTML]{333333} 118.00}       & {\color[HTML]{333333} 44.50}                                                   & {\color[HTML]{333333} 22.00}                                                  &
		{\color[HTML]{333333} 9.00}&

		{\color[HTML]{333333} 4.00}\\
		\rowcolor[HTML]{C0C0C0} 
		\cellcolor[HTML]{656565}{\color[HTML]{FFFFFF} 75\%}  & {\color[HTML]{333333} 288.00}       & {\color[HTML]{333333} 93.50}                                                   & {\color[HTML]{333333} 45.00}                                                  &
		{\color[HTML]{333333} 16.00}&
		{\color[HTML]{333333} 5.00}\\
		\rowcolor[HTML]{FFFFFF} 
		\cellcolor[HTML]{656565}{\color[HTML]{FFFFFF} max}  & {\color[HTML]{333333} 2837.00}       & {\color[HTML]{333333} 622.50}                                                   & {\color[HTML]{333333} 273.00}                                                  &
		{\color[HTML]{333333} 118.00}&
		{\color[HTML]{333333} 73.00}                                 
	
	\end{tabular}
	\caption{Statistics of advertisers that reach a user in different time periods: session, day, week, month and general that refers to the whole duration of the data collection.}
	\label{table:2_numberadvertisers}
\end{table}

Previous section has analyzed the exposure of FB users to ads without considering who were the advertisers behind those ads. In this section, we analyze how users are exposed to advertisers. 

\subsection{Quantification of the number of advertisers reaching FB users}
The goal is  to quantify how many advertisers reach FB users for different time windows. To obtain this information for each user we compute the median value of the number of advertisers reaching them in each of the considered time windows. Then, we apply standard statistics across users' median values to describe the results: mean, standard deviation (std), minimum, maximum, \nth{25} percentile, \nth{50} percentile (or median) and \nth{75} percentile. We compute those statistics for the following time windows: session, day, week and month. Table \ref{table:2_numberadvertisers} shows the results for each of the statistics and time-window. 

In average, a FB user is targeted by 4.5 advertisers in a session, 12 advertisers per day, 33 advertisers per week and 67 advertisers per month. If we consider the median these numbers reduce to 4, 9, 22 and 44, respectively. If we focus on the user receiving ads from more advertisers (see max row in the table) we observe an exposure to a huge number of advertisers such as 622 per month and more than 2.8K advertisers in the analyzed period. 

\subsection{Distribution of ads across advertisers}
The number of advertisers does not grow linearly with the time window used. For instance, if users are targeted in average by 12 advertisers per day a linear growth would suggest that they should be targeted by 7$\times$ more advertisers in a week, i.e., 84 advertisers. Hence, the results in table \ref{table:2_numberadvertisers} suggest that users receive ads from the same advertiser multiple times during a week or a month. Note that these ads could refer to the same campaign (i.e., the same ad delivered to the user multiple times) or different campaigns (i.e., different ads delivered from the same advertiser). 

\begin{figure}[t]
	\centering
	\includegraphics[width=\columnwidth]{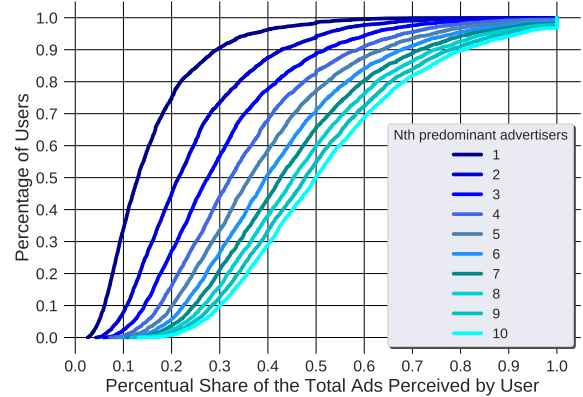}
	\caption{CDF showing the portion of ads that the N predominant advertisers represent over the total ads delivered. N ranges between 1 and 10.}
	\label{fig:AdverstiserCDF}
\end{figure}

One way to verify our hypothesis is to demonstrate that the number of ads delivered per advertiser does not follow a homogeneous distribution. To that end we compute the share of ads that the N predominant advertisers are responsible for across users. 

Figure \ref{fig:AdverstiserCDF} represents a CDF showing the accumulated share of ads delivered for the N predominant advertisers for a user with N ranging from 1 to 10. For instance the point {x=0.5, y=0.5} which appears in the line associated to N=10 shows that in median (i.e., for half of the users) the top 10 advertisers are responsible for 50\% of the ads delivered to FB users. The results show that in median a FB user receives the 13\% of the ads from the most frequent advertiser. This value increases to 21.43\% and 27.55\% when we include the second and third most predominant advertisers. As we already mentioned to explain the content of the figure, in median only 10 advertisers are responsible for half of the ads a FB user receives.

\begin{figure}[t]
	\centering
	\includegraphics[width=\columnwidth]{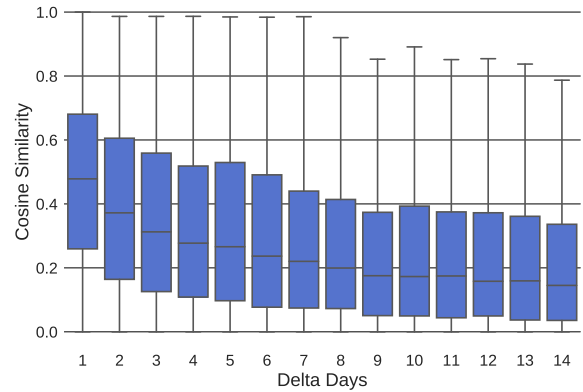}
	\caption{Cosine similarity between the pool of advertisers that delivered ads to users the first and nth consecutive days within a 15 days interval.}
	\label{fig:AdveritsersCos}
\end{figure}

\subsection{Temporal analysis of the pool of advertisers using similarity}
We zoomed in our analysis to try to verify whether in short periods of time (e.g., two weeks) the pool of advertisers targeting FB users remains stable or not. We analyze how similar the pool of advertisers reaching a user a given day (referred to as as \textit{day 0}) is to the pool of advertisers that reach that same user the day after (referred to as as \textit{day1}), two days after (\textit{day 2}),..., until two weeks later (\textit{day 14}). To compute the similarity among the pool of advertisers we use the cosine similarity, which in this case is bounded between 0 (the pool of advertisers in two different days is completely different) and 1 (the pool of advertisers in two different days is the same).

Figure \ref{fig:AdveritsersCos} shows the cosine similarity evolution from \textit{day 1} to \textit{day 14} represented with one boxplot per day. The boxplot considers all FDVT users and represents the distribution of the similarity of the pool of advertisers across the FDVT users.  There are two main outcomes from the results: (i) as expected, the pool of advertisers changes over time. This means that every day new advertisers arrive to the pool reducing the similarity with the initial pool of advertiser. (ii) After two weeks we can still find advertisers in the pool that were present at \textit{day 0}. This means that some advertisers target users over long periods of time to engage them. We further analyze in Section \ref{sec:engagement} the success probability of advertisers willing to engage users targeting them multiple times.

\subsection{Summary of results}
In summary, the main outcomes of this section are: (i) FB users are exposed to a large number of advertisers from which few of them are responsible for an important share of the ads delivered to the user. (ii) Some advertisers target users over rather long and continuous periods of time, e.g. two weeks. 
\section{Users' profiling analysis}

\begin{figure}[t]
	\centering
	\includegraphics[width=\columnwidth]{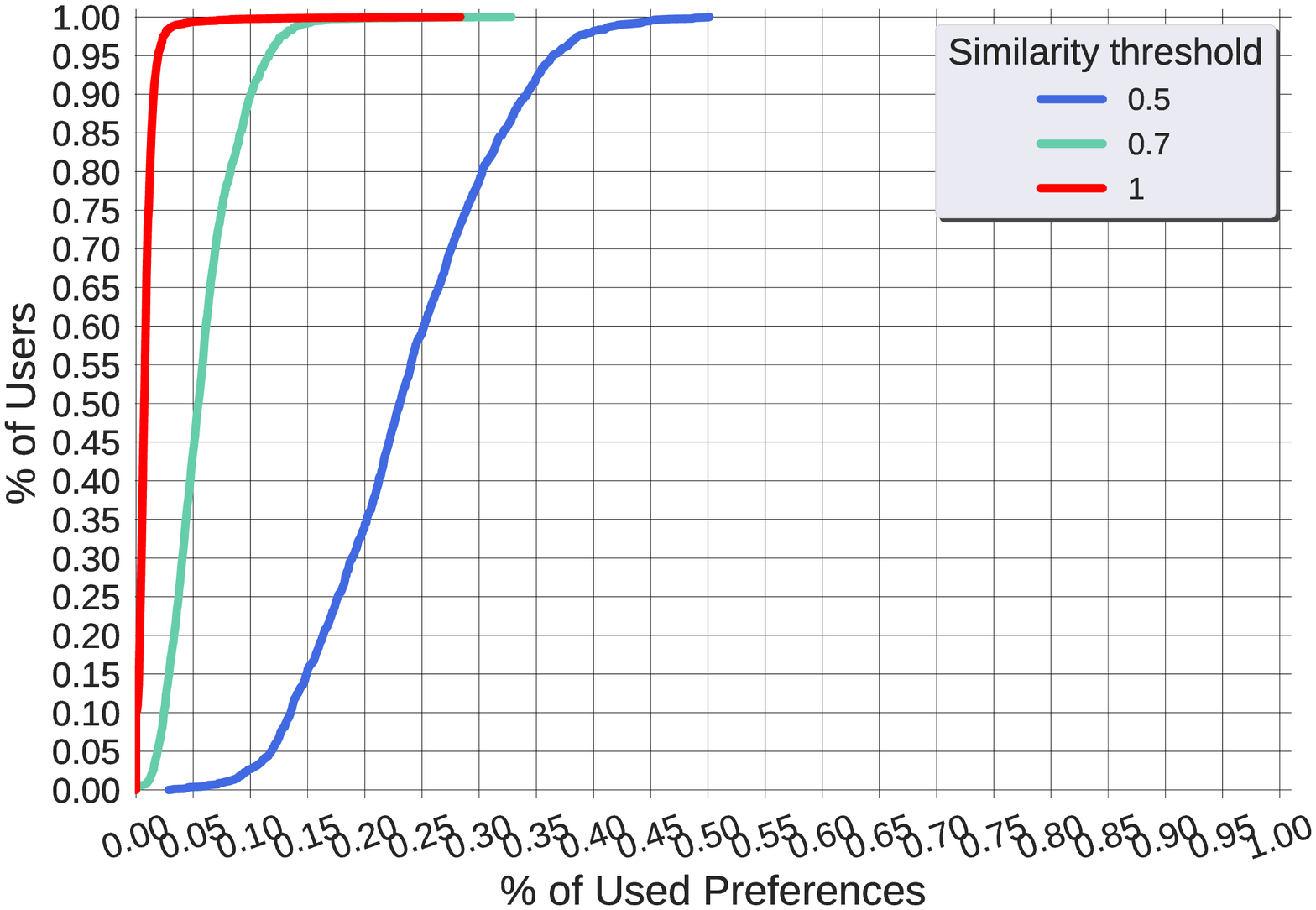}
	\caption{CDF showing the portion of users (y axis) for which X\% of the assigned FB ad preferences are used (x axis). We consider that an ad preference is used when it has a similarity higher or equal than a given similarity score. We have used three similarity scores in the figure  0.5 (blue line), 0.7 (green line) and 1 (red line).}
	\label{fig:match}
\end{figure}    

One of the most important tasks within the online advertising ecosystem is to profile users inferring their interests out of their online activity. For instance, FB infers users' interests (also known as \textit{ad preferences}) from the activity of the users in FB: pages liked by the users, ads clicked, etc. The final goal of profiling a user is delivering ads associated to her interests in order to maximize the probability of engaging the user (e.g., click on the ad). 

In this section we study whether FB is \textit{overprofiling} users or not. By \textit{overprofiling} we mean FB is tagging users with a large amount of interests from which most of them are actually unrelated to the ads users receive. Overprofiling a user may have multiple negative effects. First, creating very wide user profiles increases the probability of labeling users with interests that may compromise the privacy of users. We demonstrated this in a previous work that analyzes the number of users in FB that are assigned potentially sensitive interests \cite{Jose_usenix}.  Second, if users are overprofiled they are offered to advertisers as a very wide product with too many interests from which an important portion of them are very likely not to be relevant for the user. For instance, in our FDVT dataset we find 1,202 users that have been assigned more than 1,000 different FB preferences since they were FDVT users. It is very unlikely all those interests are actually relevant for the users.  Instead, advertisers would benefit from narrow but very accurate profiles (regarding real interests of users) in order to maximize the probability of engaging the user.

To evaluate how many active FB interests of a user are related to the received ads we compute the semantic similarity between each ad's categories and the active interests of the users. In case there is one (or more  active interests) for which the similarity is high enough we consider that interest(s) as relevant. We use the following data to compare ads and interests: (i) all the ad preferences (i.e., interests) FB has assigned to a user, which are collected using the FDVT; (ii) the label(s) assigned to each ad using the categories McAfee assigns to the advertiser (i.e., landing domain) behind the ad. To compute the semantic similarity between each pair $<$ad, interest$>$ we use the methodology we defined in \cite{Jose_usenix}. This methodology provides a similarity score ranging between -1 (the two compared terms are very different from a semantic point of view, ideally antonyms) and 1 (the two compared terms are semantically very close, ideally synonyms). Finally, it is important to note that in median FDVT users are assigned 320 active ad preferences, thus we just consider FDVT users that have received at least 320 ads (i.e., 1939 users) in order to avoid reporting a biased result.  

Figure \ref{fig:match} depicts a CDF that shows the portion of users (y axis) for which a given portion of active ad preferences (x axis) are actually relevant using three different similarity thresholds: 1, 0.7 and 0.5. As we have described above we consider an active ad preference relevant if it presents a semantic similarity equal or higher than the defined threshold with at least one of the ads delivered to the user. If the similarity threshold is 1 we only consider an ad preference relevant to an ad topic if they are almost synonyms (e.g., $<$adware, malware$>$), while if the threshold is 0.7 or 0.5, they need to be semantically similar but not necessarily synonyms (e.g,. $<$art, sculpture$>$ present a semantic similarity score of 0.7, and $<$art, interior design$>$ of 0.5). 

In median (i.e., y axis = 0.5) the portion of ad preferences related to the received ads is 0.71\%, 5.38\% and 22.76\% when the similarity threshold is 1, 0.7 and 0.5, respectively. If we consider that in median FDVT users are assigned 320 ad active preferences, the obtained result demonstrates that between  247.17 (threshold 0.5) and 317.73 (threshold 1) of the assigned ad preferences are not related to any of the ads that the users receive.  

In summary, we can conclude that most FB users are actually \textit{overprofiled} since FB assigns them many ad preferences with very little chances to be targeted by advertisers.

\section{Users' engagement}
\label{sec:engagement}

\begin{figure}[t]
	\centering
	\includegraphics[width=\columnwidth]{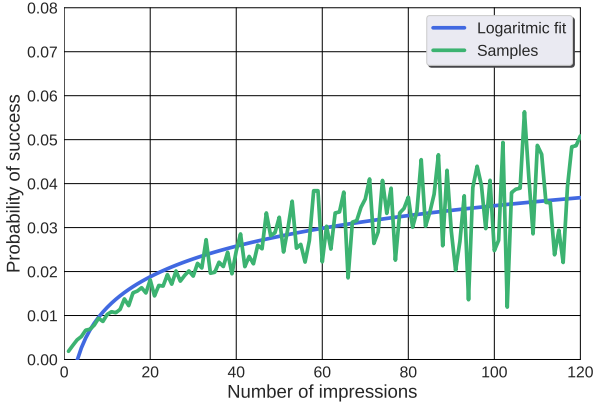}
	\caption{Probability of and advertiser of gathering at least one user click (i.e., probability of success) according to the number of impressions delivered to the user. The number of impressions ranges from 1 to 120. The figure shows the logarithmic fitting for the obtained results, which is the one minimizing the mean square error.}
	\label{fig:4_log}
\end{figure}

The ultimate goal of ad campaigns is to engage users with the product or service companies advertise. The industry uses different standard metrics to measure such engagement such us: (i) \textit{click through ratio (CTR)}, which measures the frequency of clicks on the delivered ads, (ii) \textit{conversion rate}, which measures the frequency of a particular action (e.g., purchase) derived from ads delivered to users, (iii) \textit{return of investment (ROI)}, which quantifies the benefit obtained out of an ad campaign by reducing the investment (or cost) of the campaign from the revenue obtained through the purchases derived from the campaign. 

An important metric advertisers can configure to optimize the engagement of their advertising campaigns is the so called \textit{frequency cap}.  The frequency cap establishes how many times at most the ad (from a particular campaign) can be delivered to a user. For instance, if an ad campaign defines a frequency cap of 10, no user will receive the ad associated to that campaign more than 10 times. The exposition of a user to a excessive number of impressions of the same ad may have an adverse effect, negatively affecting the opinion of the user about the product/service/brand advertised.. For instance, we can find some extreme cases in our dataset, such as the advertiser ”deliveroo.co.uk”, that delivers 6,175 impressions to a single user in 118 different days and 973 sessions. Some works in the literature have performed experiments to understand what would be a reasonable boundary for the frequency cap \cite{adscape}\cite{frequency_cap2}\cite{frequency_cap3} concluding that they should not be too high in most of the cases. However, to the best of our knowledge there is no previous analysis of frequency cap in FB. 

In this section we measure the probability of engaging FDVT users for different number of ad impressions and discuss the results in the context of the frequency cap. We use the CTR as measure of engagement  since it is the only one that can be externally measured. The conversion rate and ROI are internal metrics very rarely disclosed by advertisers or any other online advertising player. In addition, we also quantify the success probability for the first, second, third, etc. impressions to understand whether it is more likely to engage the user when she is first exposed to an ad or instead it uses to get engage after a given number of impressions.

\subsection{Measuring success probability according to the number of impressions}

We measure whether the probability of engaging a user grows with the number of ads impressions coming from a particular advertiser.  We consider that an advertiser succeeds if it is able to get at least one click of the user across N ads impressions. Figure \ref{fig:4_log} shows the success probability for the values of N (i.e., number of impressions from the same advertiser) ranging from 1 to 120.

The success probability grows logarithmically which means that there is more chances to gather a user click as the advertiser increases the number of impressions of their ads. The immediate conclusion out of this result seems to be that the higher the frequency cap the better. However, the fact the growth is logarithmic means that the increment in the success probability from N to N+1 is smaller as N increases. For instance, considering the fitting, for N=20, N=21 and N=22, the success probability is 0.0193, 0.0198 and 0.0202. The difference between N=20 and N=21 is 0.0005, whereas the difference between  N=21 and N=22 is 0.0004 (20\% lower). If we consider that: (i) the cost function per impression is linear (i.e., delivering 10 ads is 10$\times$ more expensive than delivering 1 ad), and (ii) the incremental benefit as the number of impressions increases is sub-linear (i.e., logarithmic growth),  we can deny the conclusion that the higher the frequency cap the better. This result opens the option to define an optimization problem in order to find the best frequency cap for a particular advertiser in order to maximize the benefits out of an ad campaign. However, this optimization is out of the scope of this paper since to solve such optimization problem it is required to get access to conversion rates and ROI values of advertisers. Finally, it is important to note that the optimal frequency cap  will vary from advertiser to advertiser.

\begin{figure*}[t]
	\centering
	\includegraphics[width=\textwidth]{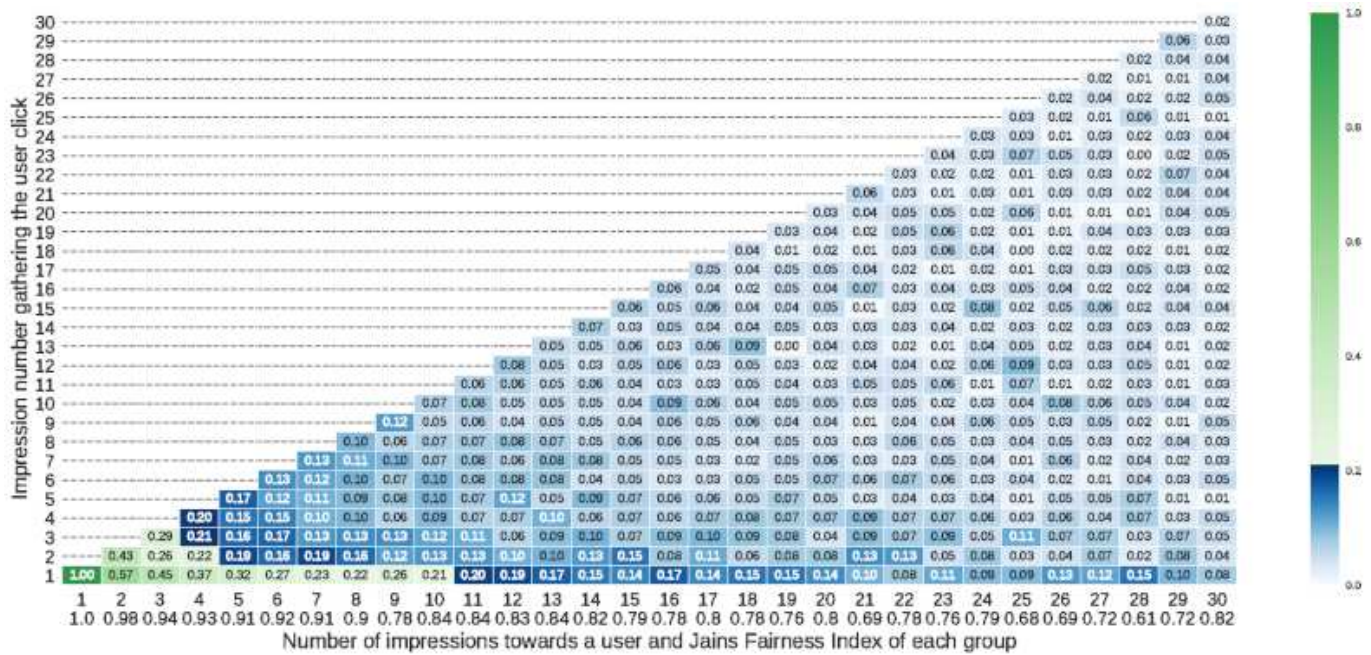}
	\caption{Probability of gathering a click in the n-th impression from ads delivered by an advertiser to a user. The number of impressions evaluated ranges from 1 to 30. The figure also shows the Jains' Fairness Index for each x axis value.}
	\label{fig:4_heat}
\end{figure*}

\subsection{Evaluating success probability of the n-th impression}

We have measured the probability that an advertiser engages a user through a click in one of its ads according to the number of impressions the user receives from that advertiser. An interesting follow up issue is to analyze whether the event of clicking the ad happens usually in the firsts, middle or lasts impressions. If users usually get engage at the very first impressions, it may suggest advertisers to reduce the frequency cap, whereas if clicks usually occur after a large number of impressions then setting up a large frequency cap may be a good option. 

We group together all the cases in our dataset in which an advertiser delivers N (ranging from 1 to 30) impressions of its ads to a user and engage the user with at least one click in one of the impressions. Following, we compute the probability that the user clicks happen in the first, second, third,..., or N-th impression. For instance, for N=10 we quantify the distribution of user clicks between the first and the tenth impression, while for N=30 we compute the distribution of user clicks between the first and the thirtieth impression. In addition, we compute the Jain Fairness Index (JFI) \cite{JainFairnessIndex} of the distribution for each value of N. The JFI ranges between 0 and 1, the closer it is to 1 the more homogeneous the distribution is. 

Figure \ref{fig:4_heat} shows a heatmap in which the x-axis represents the number of impressions (N), and the y-axis represents the impression in which the user clicks on the ad. Each point $\{x_i,y_j\}$ is represented by a box that includes the probability that in the case an advertiser delivers $i$ impressions the user clicks occurs in the $j^{\text{th}}$ impression, where $j\leq i $. In addition, each box is coloured according to the heat map scale that ranges between 0 (white) and 1 (dark green). Finally, at the bottom of the figure, below the values of the x-axis, the figure presents the JFI associated to the column (i.e., that value of N).

In most of the cases the JFI is above 0.75 which denotes that the clicks are homogeneously distributed across the impressions. However, also in most of the cases the highest click probability appears in the first impression. Only for x=15, x=21 and x=22 the first impression is not getting the highest probability. Therefore, the results suggest that the first impression is actually the one having a slightly higher engagement power. After the first impression the distribution of clicks is rather homogeneous and it is unfeasible to predict which impression will attract the click of the user.   

\subsection{Summary of results}
The main outcomes of this section are: (i) the probability of users clicking on ads from a particular advertisers grows logarithmically to the number of ads impressions received from that advertiser. The logarithmic growth denotes that for many advertisers it would be possible to obtain and optimal frequency cap to maximize their ROI; (ii) the first ad impression is the one showing a slightly higher engaging power, while all the other impressions present a homogeneous engagement probability.

\section{Related work}
\label{section:related}

The huge importance of online advertising in the sustainability of the current Internet have attracted the attention of researchers who have published a numerous number of research works (mostly in the last 5 years). These works can be divided into three major lines: optimization of online advertising from the market perspective, transparency and privacy considerations in the online advertising ecosystem and online advertising from users perspective. 

\subsection{Optimization of online advertising}

We can find a body of literature in the area of online advertising that pursuits optimizing the efficiency of digital marketing to maximize the benefits of different stakeholders. Some examples of this research line are for instance \cite{intro_perf3} and \cite{intro_perf1}.  In \cite{intro_perf3} the authors present a new scalable methodology that facilitates measuring the effects of advertising. This work proves that the adoption of the methodology leads to significant savings in the advertisers’ budget. In \cite{intro_perf1} authors evaluate the consequences that online ads  have on the advertiser’s competitors. Using data from randomized field experiments on a restaurant-search website, the authors prove that non-advertised competitors are benefited from the ads shown to the users, and that the spillover benefit decreases as the intensity of the targeting increases. Although our paper does not aim to provide any improvement from the advertiser point of view, in the user engagement analysis carried out in Section \ref{sec:engagement} we conclude that there is room to improve online advertising campaigns in FB by finding an optimal frequency cap that may help to increase the ROI of advertisers. In addition, our work reveals that FB users are overprofiled that, as we have discussed, has negative implications for the advertisers. 

\subsection{Transparency and privacy}

The main source of information for the operation of the online advertising ecosystem is data collected from end-users. Part of the collected data lays under the category of personal data and may have privacy implications. Under this context several research groups have carried out research to increase the transparency within the online advertising ecosystem and create awareness among end-users regarding how such ecosystem uses personal data to make money. Some examples of this research line are \cite{Juanmi_Conext}\cite{targetingMobile}\cite{FDVT}. In \cite{Juanmi_Conext} the authors develop artificial online entities called personas to measure volume of Online Behavioral targeted Advertising (OBA) received by users while browsing in the Internet. An interesting result of this work is that the  costlier categories in terms of CPM were more intensely targeted by advertisers. In \cite{targetingMobile},  the authors analyze the targeted advertising in the Google AdMob advertising network and extract insights about the relevance of Google user profiles, and the categories of apps used on the in-app ads served on smartphones. We have also contributed to create awareness through the development of the FDVT \cite{FDVT} that informs users of the revenue they generate for FB out of the commercial exploitation of their personal data.  

More related to our work we can find few works that analyze potential privacy issues linked to the advertising platform in Facebook. In \cite{EURECOM+5420} the authors show how Facebook third-party tracking JavaScript can be exploited by an attacker to retrieve personal data (e.g., mobile phone numbers) associated with users visiting the attacker's website. In \cite{speicherpotential} the authors demonstrate that FB ad preferences can be used to apply negative discrimination in advertising campaigns (e.g., excluding people based on their race). In \cite{Jose_usenix} we demonstrate that up to 73\% of FB users in Europe can be assigned sensitive ad preferences according to the definition of sensitive data included in the recent General Data Protection Regulation (GDPR) that applies to all European Union countries.  

Our work in the current paper also contributes to create transparency since we quantify the exposure of real users to ads and advertisers. For instance, our paper reveals that between 10\% and 15\% of the information they consume in their FB newsfeed are ads. This is a very informative result in terms of transparency for end-users but also other stakeholders in the online advertising ecosystem.

\subsection{Users' perspective}

We can find works in the literature that aims to understand the perception of end users regarding the ads they are exposed to. Many times these behavioural studies have as second goal to obtain conclusions that can be applied to improve the efficiency of the online advertising ecosystem (related to the first subsection of this Section). For instance,  in \cite{intro_userpercep3}, J.H. Schumann et al. show that appealing to reciprocity as an argument to make users accept personalized advertising  targeting is the best option in the majority of the cases. This may enable firms to increase consumers’ finisher rates by 70\%, according to the results obtained through an experiment and a field study. In \cite{intro_userpercep2}  the authors conduct a survey among 502 college-aged Facebook users in Taiwan and show that users respond to FB advertising and virtual brand communities in different ways. Users’ motivation for online social networking imply variable effects on their social media marketing responses. The paper concludes the importance to be aware of the targeted audience when creating the advertising content. These works differ from our paper since our scope is not analyzing the behaviour of users. 

More related to our work, in \cite{adscape}, the authors follow a user-centric approach by creating a crawler that replicates users browsing and reception of ads across multiple websites. The authors analyze the features, mechanisms and dynamics of display advertising on the web are analyzed. To this end the authors run several tests to characterize how the advertising was consumed by the users and how was the content of the ads related to each user's interests were performed. Similar to our approach in this paper, the authors use the domains associated to the gathered landing pages to identify advertisers. Subsequently ads are categorized by leveraging online categorization services so that they can map the similarity users' profiles and ads categories. The dataset employed in this study includes 175K ads. Although the spirit of this work is very similar to our research there are few important differences: $(i)$ the referred work focuses on websites while our research focuses on FB, $(ii)$ the referred work implements bots that replicate human browsing while our research relies on data coming from real users, $(iii)$ the dataset we used in our analysis is almost two orders of magnitude larger.

In a nutshell, to the best of our knowledge our work is: $(i)$ the first one that the exposure of FB users to the online advertising ecosystem; $(ii)$ the work using the largest dataset to analyze the exposure of user to online advertising.

\section{Conclusion}

To the best of our knowledge this paper is the first large-scale study that analyses the online advertising in Facebook from an end-user perspective. The study lays in a dataset that includes more than 7M ads from 140K unique advertisers delivered to more than 5K users between October 2016 and May 2018. We have analyzed four fundamental dimensions from an end user perspective: user exposure to ads, user exposure to advertisers, user profiling and user engagement. The first outcome of the paper is that in median FB users are exposed to 70 ads and 22 advertisers every week. These numbers are aligned to the fact that generally 10 advertisers were responsible for half of the ads received by the users in our study during the time span in which we collected the data. We have also demonstrated that ads represent between 10\% and 15\% of the information delivered to users in their FB newsfeeds. However, not all the users receive the same volume of ads since 8.37\% of the users are systematically exposed to a larger portion of ads. Moreover, our work reveals that FB overprofiles users since (in the best case) only 22.76\% of the assigned ad preferences are related to the ads delivered to the users. Finally, we have shown that trying to engage the user with an unlimited number of impressions is a bad strategy since user engagement (i.e., click on the received ads) grows logarithmically with the number of impressions. Aligned to this result is interesting to note that the first impression is the one showing a slightly higher engagement probability, while such probability is homogeneously distributed among the remaining impressions.

As a future work, there is the need to investigate further on the role of FB users as products (meaning they are auctioned) in order to enhance the yet small knowledge around the online advertising ecosystem. Since we continue to gather information from FDVT users, and the variability in the conditions of the scenario (i.e., the enforcement of the EU's GDPR), in the long-term future, we will replicate the experiments in different time windows, and look for significant differences between each of the moments. 

\section*{Acknowledgment}

 J.G. Cabañas acknowledges funding from the Ministerio de Economía, Industria y Competitividad (Spain) through the project TEXEO (TEC2016-80339-R) and the Ministerio de Educación, Cultura y Deporte (Spain) through the FPU Grant (FPU16/05852). A. Cuevas acknowledges funding from the Ministerio de Economía, Industria y Competitividad (Spain) and the European Social Fund (EU) through the Ramón Y Cajal Grant (RyC-2015-17732). R. Cuevas and A. Arrate acknowledge funding from the European H2020 project SMOOTH (grant number 786741). M. Calder\'on acknowledges funding from the European H2020 project TYPES (grant number 653449).

\ifCLASSOPTIONcaptionsoff
  \newpage
\fi



%

\bibliographystyle{acm}
\bibliography{related}
\end{document}